\documentclass[letterpaper]{article} % DO NOT CHANGE THIS
\usepackage{aaai2026}  % DO NOT CHANGE THIS
\usepackage{times}  % DO NOT CHANGE THIS
\usepackage{helvet}  % DO NOT CHANGE THIS
\usepackage{courier}  % DO NOT CHANGE THIS
\usepackage[hyphens]{url}  % DO NOT CHANGE THIS
\usepackage{graphicx} % DO NOT CHANGE THIS
\usepackage{natbib}  % DO NOT CHANGE THIS AND DO NOT ADD ANY OPTIONS TO IT
\usepackage{caption} % DO NOT CHANGE THIS AND DO NOT ADD ANY OPTIONS TO IT
\usepackage{algorithm}

\usepackage[utf8]{inputenc} % allow utf-8 input
\usepackage{url}            % simple URL typesetting
\usepackage{booktabs}       % professional-quality tables
\usepackage{amsfonts}       % blackboard math symbols
\usepackage{nicefrac}       % compact symbols for 1/2, etc.
\usepackage{microtype}      % microtypography
\usepackage{xcolor}         % colors
\usepackage{multirow}
\usepackage{tabularx}       
\usepackage{graphicx}
\usepackage{lipsum}
\usepackage{marvosym}
\usepackage{xspace}
\usepackage{blindtext}
\usepackage{amsmath}
\usepackage{algpseudocode}
\usepackage{enumitem}
\usepackage{amsmath}

\usepackage{newfloat}
\usepackage{listings}
\DeclareCaptionStyle{ruled}{labelfont=normalfont,labelsep=colon,strut=off} % DO NOT CHANGE THIS
\floatstyle{ruled}
\newfloat{listing}{tb}{lst}{}
\floatname{listing}{Listing}

\title{Surf3R: Rapid Surface Reconstruction from Sparse RGB Views in Seconds}
\author{
Haodong Zhu$^{1,4*}$  \quad
Changbai Li$^{1*}$  \quad
Yangyang Ren$^{1}$ \quad
Zichao Feng$^1$ \quad
Xuhui Liu$^1$ \quad\\
Hanlin Chen$^{2\dagger}$ \quad
Xiantong Zhen$^3$ \quad
Baochang Zhang$^{1}$ \quad\\
}
\affiliations{
\textsuperscript{\rm 1}Beihang University, China \\
\textsuperscript{\rm 2}National University of Singapore, Singapore \\
\textsuperscript{\rm 3}United Imaging, China \\
\textsuperscript{\rm 4}Zhongguancun Academy, China \\
}

\usepackage{bibentry}

\usepackage{footmisc}

\begin{document}
\nocopyright 
\maketitle
\setcounter{footnote}{0}  
\footnotetext[1]{These authors contributed equally.}
\footnotetext[2]{Corresponding author and Project lead. E-mail: hanlin.chen@u.nus.edu}

\begin{abstract}
Current multi-view 3D reconstruction methods rely on accurate camera calibration and pose estimation, requiring complex and time-intensive pre-processing that hinders their practical deployment.
To address this challenge, we introduce \textbf{Surf3R}, an end-to-end feedforward approach that reconstructs 3D surfaces from sparse views without
estimating camera poses and completes an entire scene in \textbf{under 10 seconds}. 
Our method employs a multi-branch and multi-view decoding architecture in which multiple reference views jointly guide the reconstruction process.
Through the proposed branch-wise processing, cross-view attention, and inter-branch fusion, the model effectively captures complementary geometric cues without requiring camera calibration.
Moreover, we introduce a D-Normal regularizer based on an explicit 3D Gaussian representation for surface reconstruction. It couples surface normals with other geometric parameters to jointly optimize the 3D geometry, significantly improving 3D consistency and surface detail accuracy.
Experimental results demonstrate that \textbf{Surf3R} achieves state-of-the-art performance on multiple surface reconstruction metrics on ScanNet++ and Replica datasets, exhibiting excellent generalization and efficiency.

\end{abstract}

\section{Introduction}
\label{sec:intro}
3D surface reconstruction is a long-standing problem that aims to create 3D surfaces of an object or scene captured from multiple viewpoints~\cite{space, space_1, voxel}.
This technique has wide applications in robotics, graphics, virtual reality, and other fields.
Traditional 3D surface reconstruction methods typically include two main approaches: Structure-from-Motion (SfM) combined with Multi-View Stereo (MVS)~\cite{ neusg, 2dgs, PGSR} or volumetric methods~\cite{atlas, NeuralRecon}. 
% The SfM+MVS pipeline involves first estimating camera poses and generating sparse 3D point clouds using SfM \cite{sfmwm, pixel, stfmr}, followed by computing per-view depth maps and fusing them into the final 3D surface using MVS techniques \cite{2dgs, sugar, PGSR}. 
The SfM+MVS pipeline involves first estimating camera poses and generating sparse 3D point clouds using SfM \cite{sfmwm, pixel, stfmr}. This is followed by computing per-view depth maps and fusing them into the final 3D surface using MVS techniques \cite{2dgs, sugar, PGSR}.
On the other hand, volumetric methods, such as Atlas \cite{atlas} and NeuralRecon \cite{NeuralRecon}, predict 3D volumes like Truncated Signed Distance Function (TSDF) from multiple views, often avoiding the explicit depth map computation. 
%Traditional 3D surface reconstruction methods typically combine Structure-from-Motion (SfM) and Multi-View Stereo (MVS), as shown in Fig. \ref{fig:head}. 
%In the first stage, SfM algorithms estimate the camera poses and often a sparse 3D point cloud from multiple images \cite{sfmwm, pixel, stfmr}. 
%In the second stage, dense MVS methods compute per-view depth maps to gnerate the final 3D surfaces \cite{NeuralRecon, atlas, 2dgs, sugar, PGSR}.
% , which is often computationally expensive, typically requiring 1–2 hours per scene on a modern GPU setup
Although these two kinds of methods achieve high-quality surface reconstruction, they often rely on prior knowledge or require nontrivial pre-processing steps, such as SfM to estimate camera intrinsics and extrinsics. These pre-processing steps often require heavy GPU computation and are time-consuming (typically taking 1–2 hours per scene on a modern GPU), making real-time inference challenging and reducing their practical usability.
% Moreover, many of them are based on per-scene iterative refinement, resulting in substantial training time and limited scalability.

\begin{figure}[!t]
  \centering
 \includegraphics[width=0.47\textwidth]{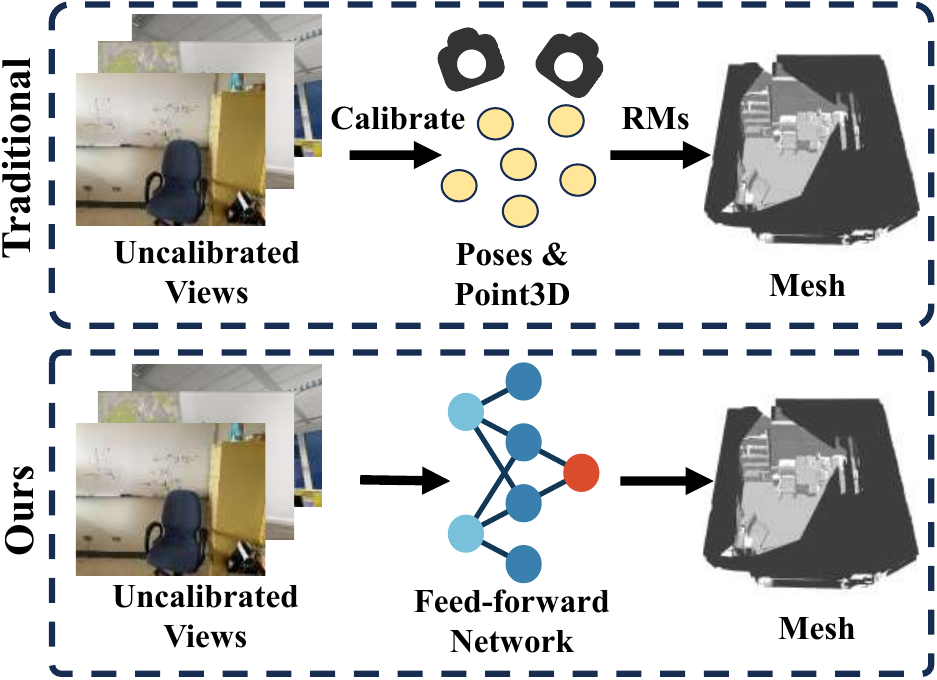}
  \caption{
  Comparison Between Traditional Methods and Our Approach.
  Traditional methods rely on SfM for sparse point clouds and calibrated poses, followed by different 3D Reconstruction Methods (RMs). 
  In contrast, our method directly reconstructs the scene from uncalibrated images in \textbf{under 10 s}, eliminating the need for calibration or iterative refinement.
  }
  \label{fig:head}
\end{figure}

To address the aforementioned limitations, inspired by DUSt3R~\cite{Dust3r}, we propose \textbf{Surf3R}, the first feed-forward network that performs pose-free surface reconstruction from sparse RGB inputs in a single pass.
Specifically, we first encode all input views using a shared encoder to extract multi-scale visual features.
To effectively model cross-view information interactions, we introduce Feature-Refine (FR) blocks that jointly learn not only the pairwise relationships between a selected reference view and all other source views, but also the interactions among source views themselves.
When reconstructing a large scene from sparse multi-view images, the geometric correspondence between a selected reference view and certain source views could be insufficient.
This is because substantial changes in camera poses make it difficult to directly infer the relation between the reference view and those source views.
To mitigate this issue, we further introduce a cross-reference fusion mechanism, implemented via a multi-branch design where multiple reference views are independently selected.
Each branch processes the input views through its own FR blocks and integrates information using dedicated Cross-Reference Fusion (CRF) blocks, enabling effective propagation of long-range and complementary information across views.
Based on the fused multi-view features, we first generate a sparse 3D point cloud for reconstruction. 
While directly converting this point cloud into a mesh using NKSR~\cite{NKSR} is feasible, our experiments (see Sec.~\ref{4.2}) show that this naive approach yields poor reconstruction quality. 
The underlying limitation is that point-cloud supervision is applied in a view-separated manner and thus lacks global 3D consistency.
We therefore adopt a Gaussian representation: each Gaussian resides in a unified 3D space and is projected into every view during rendering, so the per-view loss implicitly regularizes the entire scene and yields smoother, more accurate surfaces.
The final per-pixel Gaussian primitives are derived from specifically designed Gaussian heads.

To further facilitate accurate surface reconstruction from the predicted Gaussian parameters, we introduce a Depth-Normal Regularization strategy~\cite{VCR} designed to enhance the geometric fidelity of the reconstructed surfaces.
Specifically, we first apply a flattening operation to the Gaussian primitives to better align them with the local surface geometry.
%Inspired by prior works on depth and normal estimation~\cite{ne, dp}, 
Subsequently, we introduce a D-Normal formulation, in which surface normals are not directly blended from 3D Gaussians, but instead derived from the gradient of the rendered depth map.
% Unlike existing methods that estimate depth using the center positions of the 3D Gaussians, our approach computes depth as the intersection between the camera ray and the flattened Gaussian surfaces.
% This approach simplifies the depth computation to a ray-plane intersection, enabling a more geometrically accurate and interpretable representation.
As a result, it allows the Gaussian parameters to be directly supervised by surface normals, jointly optimizing the geometry and markedly enhancing 3D consistency and surface detail.
%Extensive experiments on the ScanNet++ and Replica datasets demonstrate that Surf3R achieves state-of-the-art surface reconstruction performance, significantly outperforming both optimization-heavy and feed-forward baselines, while also exhibiting strong generalization to unseen scenes and competitive novel view synthesis results.
Extensive experiments on the ScanNet++ datasets demonstrate that Surf3R achieves state-of-the-art surface reconstruction performance.
It significantly outperforms both optimization-based and feed-forward baselines in terms of accuracy and completeness.
%Moreover, our method exhibits strong generalization to unseen scenes 我们直接在replica上测试，few shot，展示了泛化能力。
Moreover, when evaluated on the unseen Replica dataset in a zero-shot setting, our model maintains competitive accuracy, demonstrating robust generalization to novel scenes.
And it also delivers strong performance on novel view synthesis tasks.
Notably, \textbf{Surf3R reconstructs an entire scene in under 10 seconds}, making it highly efficient for real-time or interactive applications.
We summarize our contributions as follows: 

\begin{itemize}[leftmargin=*]

\item We present \textbf{Surf3R}, the first feed-forward network for pose-free surface reconstruction from sparse multi-view RGB inputs. 
It achieves real-time surface reconstruction in \textbf{under 10 seconds}, offering both high efficiency and scalability.

\item We employ a multi-branch architecture where multiple reference views are jointly leveraged to capture long-range cross-view interactions. 
Furthermore, we introduce a Depth-Normal Regularization strategy to enhance geometric fidelity. 

\item Extensive experiments on ScanNet++ and Replica datasets demonstrate that Surf3R achieves state-of-the-art surface reconstruction, generalizes zero-shot to new scenes, and remains competitive for novel-view synthesis.
%Extensive experiments on the ScanNet++ and Replica datasets demonstrate that Surf3R not only achieves state-of-the-art performance in surface reconstruction, but also generalizes well to unseen scenes in a zero-shot setting. It also produces competitive results in novel view synthesis tasks.

\end{itemize}

%\definecolor{lightblue}{rgb}{0.7, 0.87, 0.92} 
\begin{figure*}[!t]
  \centering
  \includegraphics[width=1.0\textwidth]{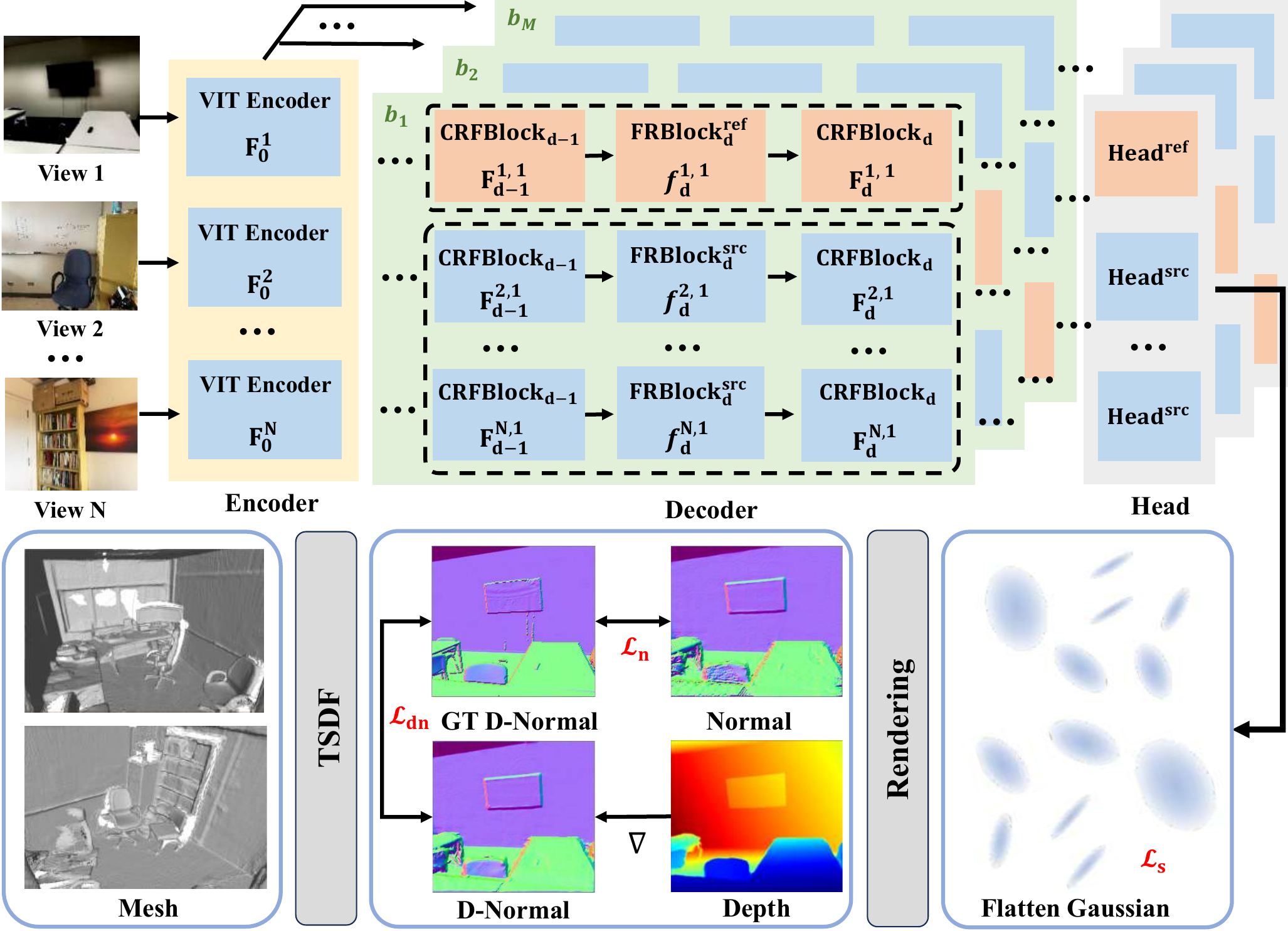}
  \caption{
  Overall Framework of \textbf{Surf3R}. 
  The reference view branches are shown in orange, while the branches of other source views are shown in blue. 
  Each model path uses a different reference view. For clarity, only one of the stacked \textbf{FRBlock} and \textbf{CRFBlock} is displayed.
  }
  \label{fig:surf3r}
\end{figure*}

\section{Related Works}

\noindent\textbf{Multi-View Surface Reconstruction.} 
Multi-view surface reconstruction recovers dense geometry from images captured at multiple viewpoints. Classical pipelines fuse depth maps obtained by multi-view stereo \cite{mvs,seitz2006comparison,schonberger2016pixelwise,zhang2020visibility,yao2018mvsnet,bleyer2011patchmatch} or optimize voxel occupancy fields \cite{de1999poxels,kutulakos2000theory,space,voxel}, but these approaches are limited by memory and cross-view noise \cite{tulsiani2017multi,ummenhofer2017demon}. Implicit neural representations such as signed distance fields \cite{Deepsdf,dist,ma2023towards,metasdf} and neural volume rendering \cite{yariv2021volume,wang2021neus} alleviate some constraints, yet methods like NeuS \cite{wang2021neus}, MonoSDF \cite{monosdf}, and Geo-NeuS \cite{Geoneus} remain optimization intensive and do not scale well. 3D Gaussian Splatting (3DGS) \cite{kerbl3Dgaussians} introduces an explicit alternative that rasterizes anisotropic Gaussians in real time \cite{yifan2019differentiable}, though the vanilla version lacks geometric supervision. Extensions such as VastGaussian \cite{lin2024vastgaussian}, SuGaR \cite{guedon2023sugar} and HRGS \cite{HRGS} incorporate view-consistent depth and normal constraints, substantially improving accuracy and convergence. 
However, these methods often require pre-processing step, limiting their practical deployment.
To bridge this gap, we present a feedforward network that dispenses with costly pre-processing while reconstructs high idelity 3D susrface from sparse views in under 10 seconds.

\noindent\textbf{Novel View Synthesis.} 
% Novel View Synthesis (NVS) focuses on generating realistic images of a scene from unseen viewpoints. Earlier approaches such as Soft3D~\cite{penner2017soft} and voxel coloring~\cite{voxel} introduced geometry-aware rendering, while NeRF~\cite{Nerf} and Mip-NeRF~\cite{barron2021mip,barron2022mip} pioneered the use of neural radiance fields for high-fidelity view synthesis via volumetric rendering. However, NeRF-based methods are often slow to train and infer due to the need for dense ray sampling~\cite{reiser2021kilonerf}.
% To accelerate optimization, methods such as InstantNGP~\cite{muller2022instant}, Plenoxels~\cite{fridovich2022plenoxels}, and KiloNeRF~\cite{reiser2021kilonerf} propose hash encoding or voxel-based acceleration structures, significantly improving training efficiency. Yet, they often compromise on scalability or struggle with sparse-view setups~\cite{garbin2021fastnerf,chen2022tensorf}.
% Recent progress in unstructured scene representations centers around 3D Gaussian Splatting~\cite{kerbl3Dgaussians}, which enables real-time rendering by directly projecting Gaussian ellipsoids onto the image plane.
% Extensions such as 2D Gaussian Splatting~\cite{2dgs}, DN-splatter~\cite{turkulainen2024dnsplatter}, Neusg~\cite{neusg}, and Gnesf~\cite{chen2023gnesf} integrate depth-normal priors or semantic constraints for improved consistency across views. Additionally, generative approaches like Dreamgaussian~\cite{twe} and Lightgaussian~\cite{fan2023lightgaussian} further expand the application of Gaussians into synthesis and editing.
Novel-view synthesis has progressed from geometry-aware volumetric grids (e.g., Soft3D \cite{penner2017soft}, voxel colouring \cite{voxel}) to neural radiance fields such as NeRF \cite{Nerf} and Mip-NeRF \cite{barron2021mip}, which deliver high fidelity but are hampered by dense ray sampling. 
Hash-encoded feature grids and sparse voxel accelerations (InstantNGP \cite{muller2022instant}, Plenoxels \cite{fridovich2022plenoxels}, KiloNeRF \cite{reiser2021kilonerf}) mitigate runtime yet remain limited in sparse-view or large-scale scenarios \cite{garbin2021fastnerf,chen2022tensorf}. Recent work on 3D Gaussian Splatting (3DGS) \cite{kerbl3Dgaussians} achieves real-time rendering by rasterising anisotropic Gaussians, achieving faster and higher-quality rendering without neural components. In this work, we utilize the advantage of Gaussian Splatting to perform surface reconstruction and incorporate normal priors to guide the reconstruction.

\noindent\textbf{Learning-based 3D Reconstruction.} 
Learning-based 3D reconstruction approaches have recently seen a lot of progress.
Notably, DUSt3R \cite{Dust3r} generates two point maps for an input image pair within a shared coordinate system, implicitly incorporating both intrinsic and extrinsic camera parameters. Nevertheless, the method is intrinsically pair-wise and lacks native support for multi-view input. Multi-view operation is possible only by appending a separate global-alignment stage that registers the individual reconstructions. This alignment must be performed offline, incurs substantial computational overhead, and cannot adapt on-the-fly when additional views become available.
Several recent studies \cite{MASt3R, Spann3R, flare, vggt} replace DUSt3R’s test-time optimization with feed-forward neural networks to accelerate inference, yet their primary focus lies in view synthesis and related tasks rather than high-fidelity surface reconstruction.
In this work, we close this gap by introducing a feed-forward, geometry-aware framework that delivers state-of-the-art surface reconstruction while retaining real-time efficiency.

% Novel-view synthesis (NVS) aims to render photorealistic images from previously unseen viewpoints. Early geometry-aware methods, such as Soft3D \cite{penner2017soft} and voxel colouring \cite{voxel}, relied on explicit volumetric grids. The introduction of neural radiance fields (NeRF) \cite{Nerf} and its anti-aliasing variant Mip-NeRF \cite{barron2021mip,barron2022mip} shifted the paradigm to volumetric neural rendering, achieving high fidelity at the cost of slow training and inference caused by dense ray sampling \cite{reiser2021kilonerf}.
% To accelerate optimisation, hash-encoded feature grids (Instant-NGP \cite{muller2022instant}) and sparse voxel structures (Plenoxels \cite{fridovich2022plenoxels}, KiloNeRF \cite{reiser2021kilonerf}) were proposed, but they compromise scalability or struggle in sparse-view settings \cite{garbin2021fastnerf,chen2022tensorf}. More recently, 3-D Gaussian Splatting (3DGS) \cite{kerbl3Dgaussians} enables real-time rendering by projecting anisotropic Gaussians directly onto the image plane. Extensions that inject geometric or semantic priors—such as 2-DGS \cite{2dgs}, DN-Splatter \cite{turkulainen2024dnsplatter}, NeuSG \cite{neusg}, and GNesf \cite{chen2023gnesf}—improve cross-view consistency and reconstruction accuracy. Beyond reconstruction, generative variants like DreamGaussian \cite{twe} and LightGaussian \cite{fan2023lightgaussian} broaden the scope of Gaussian splatting to scene synthesis and editing.

\section{Methodology}
%Surf3R pioneer in reconstructing the surface geometry of a scene from a sparse set of RGB images, without requiring known camera intrinsics or poses. Specifically, Surf3R leverages a multi-path network to parameterize 3D Gaussian primitives in a feedforward pass from multi-view inputs. To address the limitations in traditional normal regularization techniques, Surf3R introduces a Depth-Normal regularization strategy for full optimization of the Gaussian geometric parameters by combining depth and normal information. This involves formulating planar normals and intersection depths from the predicted 3D Gaussian primitives, which are used to render detailed normal maps and depth maps. Subsequently, the rendered maps are employed to enhance the regularization process, ensuring a more precise and consistent alignment of the Gaussians with the true scene geometry.

Our proposed \textbf{Surf3R} framework enables accurate surface reconstruction from a sparse set of RGB images, without the need for known camera intrinsics or poses. First, we introduce a feedforward network that extracts multi-view features and predicts per-pixel 3D Gaussian parameters (Sec.\ref{3.1}). These Gaussians are then flattened into 2D planes to better represent the surface, which helps achieve more accurate depth estimation for reconstruction (Sec.\ref{3.2}). Finally, the model is optimized using geometry-aware loss functions (Sec.~\ref{3.3}).

%
% Our proposed \textbf{Surf3R} framework effectively reconstructs surface geometry from a sparse set of RGB images, without relying on known camera intrinsics or poses. 
% We first introduce a feedforward network that extracts multi-view features and regresses per-pixel 3D Gaussian parameters (Sec.~\ref{3.1}). Then, we flatten the Gaussians into local planes to derive more accurate depth and surface representations (Sec.~\ref{3.2}). Finally, we optimize the reconstruction using a combination of geometry-aware losses (Sec.~\ref{3.3}).

\subsection{Feedforward Geometry Reconstruction}\label{3.1}
As shown in Fig.~\ref{fig:surf3r}, Surf3R employs a feedforward network architecture to extract and fuse visual information across multiple input views without camera intrinsics and poses. Unlike traditional methods, which depend on estimating these parameters, Surf3R employs a single feedforward pass to directly reconstruct 3D surfaces through view-specific processing.

\noindent\textbf{Multi-view Feature Encoding and Fusion.} 
Given $N$ input images $\{I_i\}_{i=1}^N$, a shared-weight Vision Transformer (ViT) is first employed to encode each image $I_i$ into visual tokens $F_0^i = \operatorname{ViT}(I_i)$, with a spatial resolution reduced by a factor of 16.
To mitigate the limitation that a single reference view may not provide uniformly accurate geometric cues across the entire scene, we introduce a multi-branch architecture in which multiple reference views collaboratively participate in the reconstruction process.
Specifically, we select $M$ reference views $\{r_m\}_{m=1}^M$ to construct $M$ decoding branches $\{b_m\}_{m=1}^M$, each centered around a different reference view. 
Within each branch $b_m$, a dedicated multi-view decoder, consisting of $D$ stacked Feature-Refine (FR) blocks, is employed to refine features via cross-view attention. These blocks are denoted as $\text{FRBlock}^{\text{ref}}_d$ for the reference view $r_m$ and $\text{FRBlock}^{\text{src}}_d$ for the remaining $N{-}1$ source views, where $d \in \{1, \ldots, D\}$.
At each decoding layer $d$, the token representations are updated in a view-specific manner. For a given view $I_v$, the decoder block takes as input the primary tokens $F^{v,m}_{d-1}$ from view $I_v$ and the secondary tokens $\mathcal{F}^{-v, m}_{d-1} = \{F^{i, m}_{d-1} \mid i \neq v\}$ from all other views. The update is performed as:
\begin{equation}
\small
f^{v, m}_d = 
\begin{cases}
\text{FRBlock}^{\text{ref}}_d(F^{v, m}_{d-1}, \mathcal{F}^{-v, m}_{d-1}), & \text{if } v = r_m \\
\text{FRBlock}^{\text{src}}_d(F^{v, m}_{d-1}, \mathcal{F}^{-v, m}_{d-1}), & \text{otherwise}.
\end{cases}
\end{equation}
To further enhance the expressiveness of feature representations, we introduce a Cross-Reference Fusion (CRF) block after each decoder block to fuse and update per-view tokens computed under different reference views. Specifically, the updated feature is computed as:
\begin{equation}
    \small
    \mathcal{F}^{v, m}_d = \text{CRFBlock}_d(f^{v, m}_d, f^{v, -m}_d),
\end{equation}
where $f^{v,-m}_d = \{ f^{v,1}_d, \ldots, f^{v,m-1}_d, f^{v,m+1}_d, \ldots, f^{v,M}_d \}$ denotes the set of representations for view $v$ at layer $d$ across all other reference branches.

\noindent\textbf{Gaussian Parameterizing.} 
Based on the fused multi-view features, we derive a sparse 3D point cloud. 
Direct meshing with NKSR \cite{NKSR} is possible, but Sec.~\ref{4.2} shows that it yields poor surface fidelity because global 3D consistency is absent.
Accordingly, we adopt a unified 3D Gaussian representation whose projections into all views let per-view losses regularize the entire scene, producing smoother and more accurate surfaces.
To predict the final Gaussian parameters from $F^{v, m}_D$, we introduce two types of heads: $\text{Head}^{\text{ref}}$ for the reference view and $\text{Head}^{\text{src}}$ for the remaining source views. Each head comprises two sets of regression branches. 
The first set includes a pointmap head and a confidence head, which respectively predict a 3D pointmap $P^{v, m} \in \mathbb{R}^{H \times W \times 3}$ and a confidence map $C^{v, m} \in \mathbb{R}^{H \times W}$ for each view. 
The second set consists of Gaussian-specific heads that regress the per-pixel Gaussian parameters, including scaling factors $S^{v, m} \in \mathbb{R}^{H \times W \times 3}$, rotation quaternions $q^{v, m} \in \mathbb{R}^{H \times W \times 4}$, and opacity values $\alpha^{v, m} \in \mathbb{R}^{H \times W}$, which are essential for novel view synthesis.
Notably, the predicted pointmap serves as the center of the Gaussian, the input pixel color $I_v$ is used for its color and fix the spherical harmonics degree to be 0.
During inference, A model with M branches is used but the final per-view Gaussian
predictions are computed using the heads in the first branch.

%During inference, Surf3R uniformly selects $M$ reference views, always including the first view, and processes all views through its multi-path architecture. The final per-pixel Gaussian primitives are derived from the heads of the first reference path, allowing Surf3R to reconstruct the scene's surface geometry and appearance with high fidelity in a direct, feedforward manner.

\subsection{Planar Geometry Formulation}
\label{3.2}
%Prior to introducing the Depth-Normal Regularization strategy, we first derive two planar geometric properties: normal and depth from our predicted 3D Gaussian primitives, which are used to render the corresponding normal map and depth map for regularization.
%得到Gaussian参数后，为了做表面重建，我们引入了Depth-Normal Regularization strategy来得到更加精确的深度表达。该策略需要two planar geometric properties: normal and depth from our predicted 3D Gaussian primitives。
To facilitate surface reconstruction from the predicted Gaussian parameters, we introduce a Depth-Normal Regularization strategy aimed at enhancing the accuracy of depth representation. This strategy leverages two fundamental planar geometric properties: normal and depth from our predicted 3D Gaussian primitives.

\noindent\textbf{Flattening 3D Gaussians.}
%The parameterized Gaussian is fundamentally an ellipsoidal shape defined by its center ${\mathbf{p}}$, rotation, and scaling factors. 
To enhance the capacity of Gaussians in modeling surface geometry, we first apply a flattening operation to the Gaussian primitives. Inspired by~\cite{neusg},
%To convert these 3D ellipsoids into planar shapes that accurately conform to the scene surface, we employ a flattening mechanism by compressing the Gaussian along the direction of its minimum scale factor.
we specifically introduce a scale regularization loss $\mathcal{L}_s$, which minimizes the smallest of the three scaling factors $\mathbf{S}=\left(s_1, s_2, s_3\right)^{\top} \in \mathbb{R}^3$ for each Gaussian:
\begin{equation}
\small
\mathcal{L}_{\mathrm{s}}=\left\|\min \left(s_1, s_2, s_3\right)\right\|_1.
\end{equation}
By minimizing the loss, the Gaussian is driven towards a flat shape, effectively approximating a local surface plane.

\noindent\textbf{Normal Map Rendering.}
% \noindent\textbf{Normal and Intersection Depth.}
%Once a Gaussian is flattened onto a local plane,
%, its surface normal is clearly defined. 
%The normal of a Gaussian is represented by the direction aligned with its smallest principal axis of scaling \cite{neusg}. 
%The normal ${\mathbf{n}}$ is determined by the rotation quaternion $q$ and scaling factors $S$ predicted by our multi-path network: $\mathbf{n}={R}[k,:] \in \mathbb{R}^3, k=\operatorname{argmin}\left(\left[s_1, s_2, s_3\right]\right)$, These extracted normals are consistently represented within the camera coordinate system.
Once a Gaussian is flattened onto a local plane, the surface normal $\mathbf{n}$ is computed from the predicted rotation quaternion $q$ and scaling factors $S$ predicted by our feedforward network. 
We first convert $q$ into a rotation matrix $R \in \mathbb{R}^{3 \times 3}$. The normal is then defined as the direction corresponding to the smallest scaling factor: $\mathbf{n}={R}[k,:] \in \mathbb{R}^3, k=\operatorname{argmin}\left(\left[s_1, s_2, s_3\right]\right)$.
The normal $\mathbf{n}$ is subsequently transformed into the camera coordinate system.
Finally, a rendered normal map $ \hat{\mathbf{N}} $ is generated by a weighted summation of individual Gaussian normals $ \mathbf{n}_i $ and their opacities $ \alpha_i $ along each ray:
\begin{equation}
\small
\hat{\mathbf{N}} = {\sum_{i \in K} \mathbf{n}_i \alpha_i \prod_{j=1}^{i-1}(1-\alpha_j)}/{\sum_{i \in K} \alpha_i \prod_{j=1}^{i-1}(1-\alpha_j)},
\label{eqn:rendered_normal_map}
\end{equation}

\noindent\textbf{Depth Map Rendering.}
To achieve more precise and geometrically consistent depth values than simply using the Gaussian's center position \cite{twe}, we compute the depth as the intersection point of a viewing ray originating from the camera center with the plane represented by the flattened Gaussian. Formally, the intersection depth $\mathbf{d}(\mathbf{n},\mathbf{p},\mathbf{r})$ is calculated by:
\begin{equation}
\small
\mathbf{d}(\mathbf{n}, \mathbf{p}, \mathbf{r}) = \mathbf{r}_z * (\mathbf{n} \cdot \mathbf{p}) / (\mathbf{n} \cdot \mathbf{r}),
\end{equation}
where $\mathbf{r}_z$ is the z-value of the ray direction. 
%It shows that the intersection depth of a Gaussian is defined by its position and normal, offering a more accurate depth calculation.
This formulation reveals that the intersection depth of a Gaussian is jointly determined by its position and surface normal, thereby enabling a more geometrically grounded and accurate depth estimation.
Leveraging this property, a view-consistent depth map $\hat{D}$ is rendered through a weighted summation of these intersection depths $d_i$, weighted by their opacities $\alpha_i$:
\begin{equation}
\small
\hat{D} = \frac{\sum_{i \in K} d_i \alpha_i \prod_{j=1}^{i-1}(1-\alpha_j)}{\sum_{i \in K} \alpha_i \prod_{j=1}^{i-1}(1-\alpha_j)},
\label{eqn:rendered_depth}
\end{equation}
where $K$ denotes the set of Gaussians along a ray, sorted by depth. The properties form the foundation for regularization.
% subsequent

% \begin{table*}[t]
% \caption{\textbf{Quantitative comparison on the ScanNet++ dataset.} Our method achieves state-of-the-art performance across all metrics.}
% %\vspace{0.2cm}
% \centering
% \renewcommand{\arraystretch}{1.2}  % 增加行间距
% \setlength{\tabcolsep}{5pt}        % 控制列间距
% \scalebox{1.0}{
% \begin{tabular}{l|cccc|cccc}
% \hline
% \multirow{2}{*}{\textbf{Method}} & \multicolumn{4}{c|}{\textbf{Per-Scene}} & \multicolumn{4}{c}{\textbf{Feedforward}}  \\
%                                 & NeuS & 2DGS & SuGaR & PGSR   & DUSt3R & Mv-DUSt3R & Mv-DUSt3R+ & \textbf{Ours} \\
% \hline
% Precision $\uparrow$   & 29.42 & 23.01 & 38.30 & 35.33 &  4.62& \textbf{\textcolor{cyan}{63.75}} & \textbf{\textcolor{red}{78.50}} & \textbf{\textcolor{red}{80.24}} \\
% Recall $\uparrow$      & 22.14 & 16.04 & 34.92 & 21.70  & 4.84& \textbf{\textcolor{cyan}{62.36}} & \textbf{\textcolor{red}{75.34}} & \textbf{\textcolor{red}{77.55}} \\
% F1-score $\uparrow$    & 25.13 & 18.30 & 36.12 & 24.92 & 4.06& \textbf{\textcolor{cyan}{62.89}} & \textbf{\textcolor{red}{76.72}} & \textbf{\textcolor{red}{78.71}} \\
% \hline
% \textit{Time}          & \multicolumn{4}{c|}{\textit{$>$ 30 min}} & $>$ 1 min &\multicolumn{3}{|c}{\textit{$<$ 10 s}} \\
% \hline
% \end{tabular}
% }
% \label{tab:scannetpp}
% %\vspace{-0.3cm}
% \end{table*}
\begin{table*}[t]
\centering
\renewcommand{\arraystretch}{1.2}  % 增加行间距
\setlength{\tabcolsep}{5pt}        % 控制列间距
\scalebox{0.9}{
\begin{tabular}{l|cccc|cccc}
\hline
\multirow{2}{*}{\textbf{Method}} & \multicolumn{4}{c|}{\textbf{Per-Scene}} & \multicolumn{4}{c}{\textbf{Feedforward}}  \\
                                & NeuS & 2DGS & SuGaR & PGSR   & DUSt3R & Surf3R-P & Surf3R-G & Surf3R-GD (Ours)\\
\hline
Precision $\uparrow$   & 29.42 & 23.01 & 38.30 & 35.33 &  4.62 & 63.75 & 78.50 & \textbf{80.24} \\
Recall $\uparrow$      & 22.14 & 16.04 & 34.92 & 21.70 &  4.84 & 62.36 & 75.34 & \textbf{77.55} \\
F1-score $\uparrow$    & 25.13 & 18.30 & 36.12 & 24.92 &  4.06 & 62.89 & 76.72 & \textbf{78.71} \\
\hline
\textit{Time}          & \multicolumn{4}{c|}{\textit{$>$ 30 min}} & $>$ 1 min &\multicolumn{3}{|c}{\textit{$<$ 10 s}} \\
\hline
\end{tabular}
}
\caption{\textbf{Quantitative comparison on ScanNet++ dataset.} \textbf{Bold} indicates best result. Our method achieves state-of-the-art performance across all metrics.
\emph{Surf3R-P}: point-map heads, trained with $\mathcal{L}_c$; \emph{Surf3R-G}: + Gaussian heads, adds $\mathcal{L}_r$; \emph{Surf3R-GD} (Full model): + D-Normal regularization, adds $\mathcal{L}_s$, $\mathcal{L}_n$ and $\mathcal{L}_{dn}$.
}
\label{tab:scannetpp}
\end{table*}

\subsection{Geometry-aware Loss Functions}
\label{3.3}
%写成优化目标或者loss function：
% 为了训练我们的模型我们用了以下loss：
% 首先是常规的rgb loss，训练gaussian。为了优化surface，我们又用了flatten loss来压扁；之后用了normal loss；但是normal loss对于优化表面效果不够，又加了dnormal。之后。。。
% 最后，我们的总的loss是xxxx
To train our model effectively, we employ a set of loss functions tailored to guide the learning of geometry-aware representations.
We begin with a confidence-aware pointmap regression loss, denoted as $\mathcal{L}_{\text{conf}}$, which supervises the predicted 3D pointmaps $P^{k,m}$ using their associated confidence maps $Q^{k,m}$. The loss is defined as:
\begin{equation}
\small
\mathcal{L}_{c} = \sum_{k,m} \sum_{p \in P^{k,m}} Q^{k,m}_p \left\| P^{k,m}_p - \mathbf{P}_{gt,p} \right\|_1 - \beta \log Q^{k,m}_p
\label{eqn:conf_loss}
\end{equation}
where $ P^{k,m}_p $ is the predicted 3D point, $ \mathbf{P}_{gt,p} $ is the ground truth, $ Q^{k,m}_p $ is the confidence score, and $ \beta $ is a regularization parameter.
%接着是常规的RGB的渲染loss$\mathcal{L}_{\text{render}}$ \cite{}.为了优化surface我们又引入了flatten loss $\mathcal{L}_{\text{s}}$ 来将Gaussian压扁。接着我们又使用了rendered normal map loss $\mathcal{L}_n $ supervises the rendered normal map $ \hat{\mathbf{N}} $ against $ \mathbf{N}_{gt} $ using a combination of L1 and cosine losses:
% \begin{equation}
%     \mathcal{L}_n = \| \hat{\mathbf{N}} - \mathbf{N} \|_1 + (1 - \hat{\mathbf{N}} \cdot \mathbf{N}).
% \label{eq2}
% \end{equation}
% This loss ensures the rendered normals reflect the true scene geometry.
We also employ a standard RGB rendering loss $\mathcal{L}_{\text{r}}$~\cite{pixel}, which supervises the rendered image against the ground-truth RGB image to preserve photometric fidelity.

To improve surface regularity, we introduce a flattening loss $\mathcal{L}_{\text{s}}$, which encourages the predicted Gaussian primitives to lie on locally planar surfaces.
In addition, we employ a rendered normal map loss $\mathcal{L}_n$ to align the rendered normal map $\hat{\mathbf{N}}$ with a reference normal map $\mathbf{N}_{\text{gt}}$, which is computed from the ground-truth depth via finite difference-based gradients.
% Here, $\mathbf{N}_{\text{gt}}$ is derived from the ground-truth depth map via finite difference-based gradient computation.
The loss combines an $\ell_1$ term and a cosine similarity term, and is defined as:
\begin{equation}
\small
\mathcal{L}_n = | \hat{\mathbf{N}} - \mathbf{N}_{\text{gt}} |_1 + \left(1 - \hat{\mathbf{N}} \cdot \mathbf{N}_{\text{gt}} \right),
\label{eq2}
\end{equation}
where the first term enforces per-pixel accuracy, and the second term promotes angular alignment. This supervision encourages the rendered normals to more faithfully reflect the underlying scene geometry.
% \subsection{Depth-Normal Regularization}

While explicit normal regularization can effectively refine the orientation of 3D Gaussians, it has less impact on their positions. To address this limitation and ensure robust 3D surface reconstruction, we introduce a Depth-Normal (D-Normal) regularization strategy~\cite{VCR}, which enables joint optimization of both the orientation and positional accuracy of the Gaussians.
The D-Normal $\overline{\mathbf{N}}_d$ is derived from the rendered depth $\hat{D}$ by computing the cross-product of horizontal and vertical finite differences from neighboring points:
\begin{equation}
\overline{\mathbf{N}}_d = \frac{\nabla_v \mathbf{d} \times \nabla_h \mathbf{d}}{\left|\nabla_v \mathbf{d} \times \nabla_h \mathbf{d}\right|},
\label{eq5}
\end{equation}
where $\mathbf{d}$ represents the 3D coordinates of a pixel obtained via back-projection from the depth map. 
Finally, the D-Normal regularization loss $\mathcal{L}_{dn}$ is defined as:
%optimizes the orientation and position of the Gaussian primitives by aligning the derived $\bar{\mathbf{N}}_d$ with the ground truth normal map $\mathbf{N}_{gt}$:
% We then apply the D-Normal regularization like VCR-GauS \cite{}:
\begin{equation}
    \small
    \mathcal{L}_{dn} =  \left(\| \bar{\mathbf{N}}_d - \mathbf{N}_{gt} \|_1 + (1 - \bar{\mathbf{N}}_d \cdot \mathbf{N}_{gt})\right),
\label{eq8}
\end{equation}

% To reconstruct scene surfaces, 
% The rendered normal map loss $\mathcal{L}_n $ supervises the rendered normal map $ \hat{\mathbf{N}} $ against $ \mathbf{N}_{gt} $ using a combination of L1 and cosine losses:
% \begin{equation}
%     \mathcal{L}_n = \| \hat{\mathbf{N}} - \mathbf{N} \|_1 + (1 - \hat{\mathbf{N}} \cdot \mathbf{N}).
% \label{eq2}
% \end{equation}
% This loss ensures the rendered normals reflect the true scene geometry.

% Surf3R enhances pointmap accuracy with a confidence-aware loss $ \mathcal{L}_{conf} $, which optimizes the 3D pointmaps $ P^{k,m} $ using confidence maps $ Q^{k,m} $:
% \begin{equation}
% \mathcal{L}_{conf} = \sum_{k,m} \sum_{p \in P^{k,m}} Q^{k,m}_p \left\| P^{k,m}_p - \mathbf{P}_{gt,p} \right\|_1 - \beta \log Q^{k,m}_p
% \label{eqn:conf_loss}
% \end{equation}
% Here, $ P^{k,m}_p $ is the predicted 3D point, $ \mathbf{P}_{gt,p} $ is the ground truth, $ Q^{k,m}_p $ is the confidence score, and $ \beta $ is a regularization parameter.

\textbf{Overall Loss.} The final total loss $ \mathcal{L}_{total} $ combines the geometric regularization losses for Gaussian primitives and the pointmap regression loss:
\begin{equation}
\small
\mathcal{L}_{total} = \lambda_{c} \mathcal{L}_{c}  + \lambda_{r}\mathcal{L}_{r} + \lambda_{s}\mathcal{L}_{s} +  \lambda_n \mathcal{L}_n + \lambda_{dn} \mathcal{L}_{dn}
\label{eqn:total_loss}
\end{equation}
The weighting factors $ \lambda_{c} $,  $ \lambda_{r} $, $ \lambda_{s} $, $ \lambda_n $, and $ \lambda_{dm} $ balance the contributions of each loss term, ensuring a holistic optimization of the reconstructed geometry and pointmaps.

\section{Experiments}
\label{4}

% xxxxx

We begin by presenting the experimental setup in Sec.~\ref{4.1}. 
We assess the effectiveness and generalization capability of our approach for surface reconstruction in Sec.~\ref{4.2}.
We further demonstrates the novel view synthesis capability of our method in Sec.~\ref{4.3}.  
Additionally, we validate the effectiveness of the proposed techniques in Sec.~\ref{4.4}.

\subsection{Implementation Details}
\label{4.1}

We train our model on ScanNet++ dataset~\cite{scannet++}.
View sequences \(\{I_v\}_{v=1}^{N}\) are generated with an overlap–based sampler.
Starting from a random keyframe, a candidate view is appended whenever the
overlap between its point cloud and the accumulated scene cloud falls within
\(30\%\text{–}70\%\).
% For training, each scene contributes \(100\) such trajectories, each containing \(10\) views,
% providing diverse yet geometrically coherent inputs.
% For validation, on the ScanNet++ validation split we build \(1\,000\) trajectories of \(30\) views each; for surface fusion we retain the \(50\) widest–baseline views per scene to maximise spatial coverage.
For training, each scene provides \(100\) trajectories of \(10\) views, yielding diverse yet geometrically consistent inputs.  
For validation, we construct \(1\,000\) trajectories with \(30\) views on the ScanNet++ validation split and retain the \(50\) widest-baseline views per scene for surface fusion to maximise spatial coverage.
Additionally, we conduct zero-shot generalization experiments on Replica dataset~\cite{replica} to assess the cross-dataset adaptability of our model.

% To construct input trajectories $\{I_v\}_{v=1}^{N}$, we follow an overlap-based sampling strategy proposed in~\cite{MV-dust3r}. 
% Specifically, for each scene, we initialize a trajectory with a randomly selected frame and include subsequent candidate views if the overlap between the candidate point cloud and the current accumulated scene point cloud falls within a predefined range of $(t_{\text{min}}, t_{\text{max}}) = (30\%, 70\%)$. 
% For training, we sample 100 trajectories per scene, with each trajectory comprising 10 views, resulting in a diverse and geometrically consistent set of training inputs. 
% At test time, we generate 1K trajectories from ScanNet++ validation set, each containing 30 input views to support evaluation under high-coverage multi-view settings. 
% For surface reconstruction during inference, we further select 50 viewpoints with large baseline variations per scene in the validation set to enhance spatial coverage. Additionally, we conduct zero-shot generalization experiments on Replica dataset~\cite{replica} to assess the cross-dataset adaptability of our model.

% We process input views at resolution 224 × 224. We utilize 32 NVIDIA H800 GPUs for the model training.
% To initialize, MV-DUSt3R+ model weights are used.
% We use the first N = 8 views of each trajectory as input views, and randomly select $M=4$ views as the reference views.
% We train for 30 epochs using 30K trajectories per epoch, which takes 24 hours. 
% More details can be found in the appendix.
We train using 32 NVIDIA H800 GPUs, processing input views at a resolution of $224 \times 224$. 
%To initialize the model, we adopt pre-trained weights from MV-DUSt3R \cite{MV-dust3r}. 
For each training trajectory, the first $N = 8$ views are used as input, from which $M = 4$ reference views are randomly selected. The model is trained for 50 epochs, resulting in a total training time of approximately 40 hours. Additional training details are provided in the Appendix.

% \textcolor{red}{
% For MVS reconstruction evaluation, to assess the performance of each method in reconstructing scenes of variable sizes, we report results with input views ranging from 4 to 24 views. For NVS evaluation, we use the remaining 6 views as
% novel views. Below we report results on all evaluation datasets for all choices of the number of input views
% using only one MV-DUSt3R model and one MV-DUSt3R+ model.
% }

% \begin{figure*}[!t]
%     \centering
%     \includegraphics[width=1.0\linewidth]{AnonymousSubmission/LaTeX/fig/mesh.pdf}
%     % \vspace{-1em}
%     \caption{
%     Qualitative comparison of surface reconstruction results on the ScanNet++ and Replica datasets.
%     }
%     \label{fig:visual}
%     \vspace{-5pt}
% \end{figure*}

\begin{figure*}[t]
    \centering
    \includegraphics[width=0.9\linewidth]{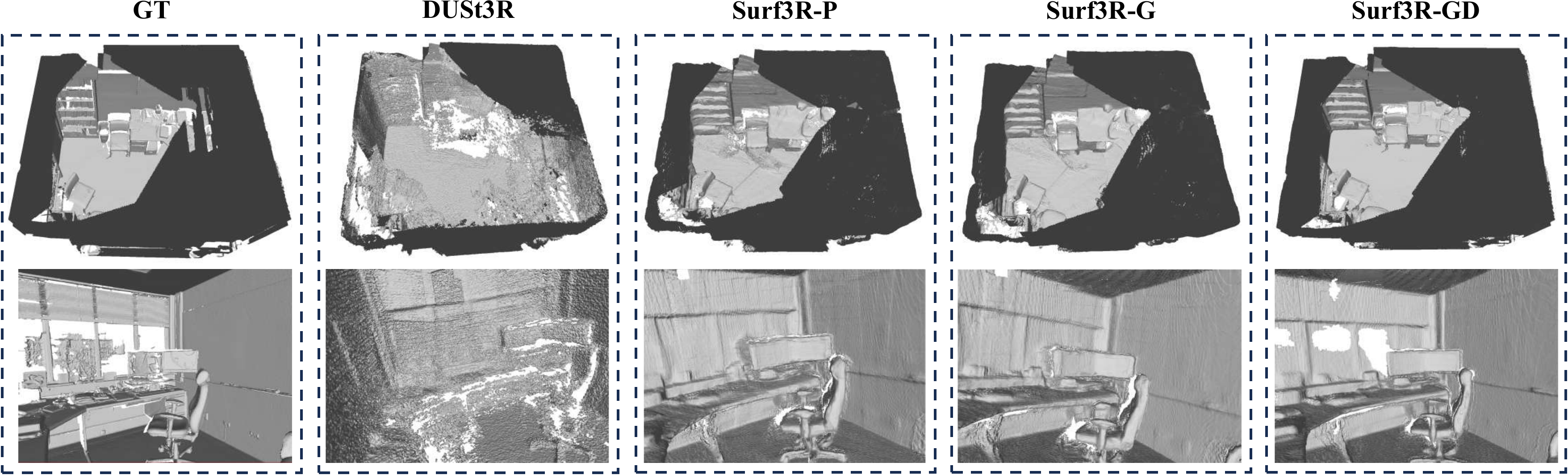}
    % \vspace{-1em}
    \caption{
    Qualitative comparison of surface reconstruction results on ScanNet++ dataset.
    }
    \label{fig:scannetpp}
\end{figure*}

\begin{figure*}[t]
    \centering
    \includegraphics[width=0.9\linewidth]{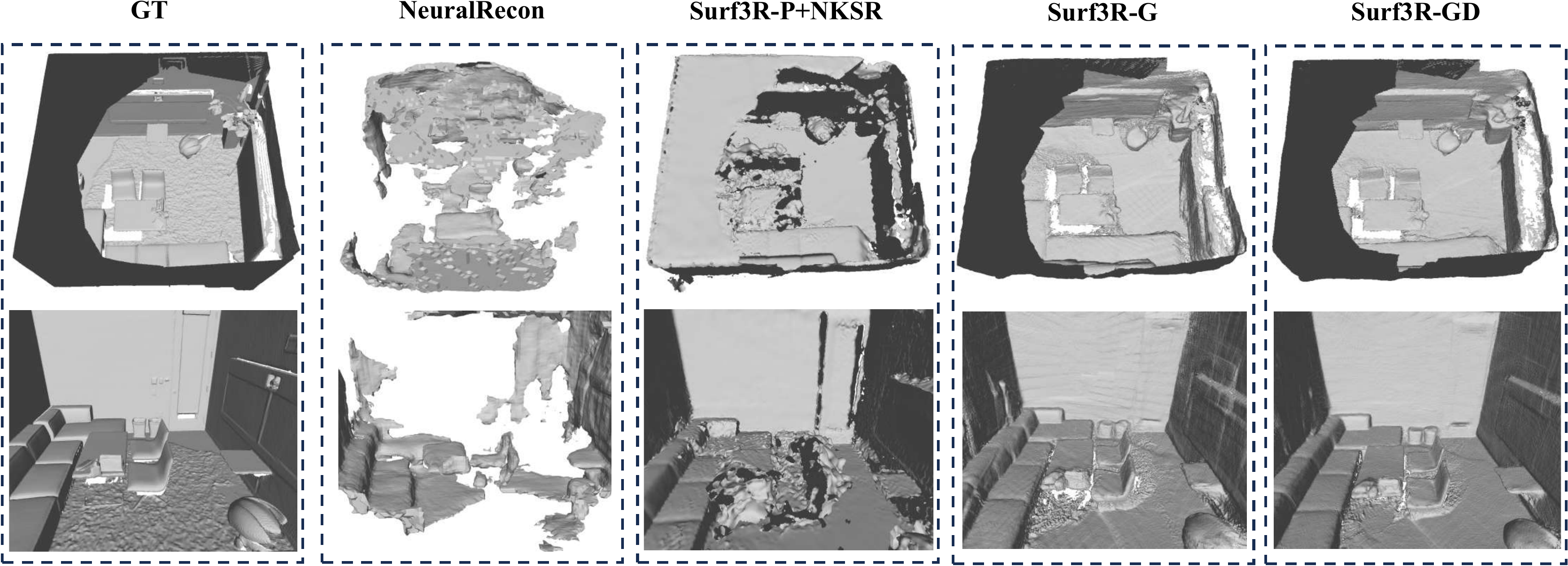}
    \caption{
    Qualitative comparison of zero-shot surface reconstruction results on Replica dataset.
    }
    \label{fig:replica}
\end{figure*}

\subsection{Surface Reconstruction} 
\label{4.2}
We evaluate three progressively enhanced variants of our framework.
\textbf{Surf3R-P} employs only the pointmap heads and is trained with the reconstruction loss $\mathcal{L}_{c}$.
\textbf{Surf3R-G} extends this baseline by introducing Gaussian heads together with the associated rendering loss $\mathcal{L}_{r}$.
\textbf{Surf3R-GD}, the full approach, further incorporates the D-Normal regularization strategy and is optimized with the additional terms $\mathcal{L}_{s}$, $\mathcal{L}_{n}$ and $\mathcal{L}_{dn}$.
As shown in Tab.~\ref{tab:scannetpp}, the results for per-scene methods are computed on the eight ScanNet++ validation scenes, while feed-forward models are evaluated across all 50 scenes, with the table reporting the dataset-wide averages. 
Surf3R-GD achieves state-of-the-art surface reconstruction performance, with an F1-score of 78.71. 
Compared to traditional per-scene reconstruction approaches such as NeuS~\cite{NeuS}, 2DGS~\cite{2dgs}, SuGaR~\cite{sugar}, and PGSR~\cite{PGSR}, our method achieves significantly higher surface reconstruction quality. 
In particular, compared to the concurrent method SuGaR~\cite{sugar}, our approach yields a substantial improvement (78.71 vs. 36.12 in F1-score). 
Moreover, our model exhibits exceptional efficiency, offering a reconstruction speed that is approximately $180\times$ faster than per-scene methods.
We further compare our method with feedforward-based approach DUSt3R~\cite{Dust3r}.
DUSt3R and Surf3R-P reconstruct surfaces by first back-projecting point clouds to depth maps, which are then fused via TSDF.
In contrast, both Surf3R-G and Surf3R-GD leverage Gaussian rendering to directly estimate high-quality depth maps for mesh reconstruction.
As shown in Tab.~\ref{tab:scannetpp}, Surf3R-P achieves a significant improvement over DUSt3R, which requires explicit global alignment, with an F1-score increasing from 4.06 to 62.89. 
This underscores the advantage of aggregating geometric cues across all input views rather than relying on pairwise stereo matches processed one at a time.
Moreover, enriching Surf3R-P with a 3D Gaussian representation (Surf3R-G) enhances global geometric consistency and raises the F1-score from 62.89 to 76.72.
And the additional introduction of D-Normal regularization (Surf3R-GD) pushes it further to 78.71, yielding the best overall performance.
As shown in Fig \ref{fig:scannetpp}, our approach yields more accurate and complete reconstructions, particularly excelling at recovering planar surfaces and capturing fine-grained geometric details.

Moreover, our method demonstrates strong generalization capabilities. As shown in Tab.~\ref{tab:replica}, under zero-shot inference on the Replica dataset, Surf3R-GD also achieves state-of-the-art performance with an F1-score of 41.92. 
Compared to traditional methods such as NeuralRecon \cite{NeuralRecon}, DUSt3R \cite{Dust3r}, and Surf3R-P+NKSR \cite{NKSR}, it consistently outperforms all baselines. As shown in Fig. \ref{fig:replica}, our method produces more complete and faithful surfaces, highlighting the superior generalization capability of our approach under unseen scenes.

\begin{table*}[ht]
\centering
\renewcommand{\arraystretch}{1.2}  % 增加行间距
\setlength{\tabcolsep}{5pt}        % 控制列间距
\scalebox{0.9}{
\begin{tabular}{l|cccccc}
\hline
\textbf{Method}  & NeuralRecon & DUSt3R & Surf3R-P & Surf3R-P + NKSR & Surf3R-G & Surf3R-GD (Ours) \\
\hline
Precision $\uparrow$    & 14.61 & 20.16 & 22.31 & 24.14 & 24.86 & \textbf{36.66} \\
Recall $\uparrow$       & 12.33 & 14.80 & 21.52 & 31.37 & 32.06 & \textbf{49.04} \\
F1-score $\uparrow$     & 13.41 & 16.90 & 21.88 & 27.16 & 27.96 & \textbf{41.92} \\
\hline
\end{tabular}
}
\caption{\textbf{Quantitative comparison on Replica dataset.} \textbf{Bold} indicates best result. Our method achieves superior performance across all metrics. \emph{Surf3R-P}: point-map heads, trained with $\mathcal{L}_c$; \emph{Surf3R-G}: + Gaussian heads, adds $\mathcal{L}_r$; \emph{Surf3R-GD} (Full model): + D-Normal regularization, adds $\mathcal{L}_s$, $\mathcal{L}_n$ and $\mathcal{L}_{dn}$.
}
\label{tab:replica}
\vspace{-7pt}
\end{table*}

\begin{table*}[t]
\centering
\renewcommand{\arraystretch}{1.2}  % 增加行间距
\setlength{\tabcolsep}{4pt}        % 控制列间距
\scalebox{0.85}{
\begin{tabular}{lccc|ccc|ccc}
\hline
   & \multicolumn{3}{c|}{\textbf{4 Views}} & \multicolumn{3}{c|}{\textbf{12 Views}} & \multicolumn{3}{c}{\textbf{24 Views}} \\ 
               & PSNR $\uparrow$ & SSIM $\uparrow$ & LPIPS $\downarrow$ 
               & PSNR $\uparrow$ & SSIM $\uparrow$ & LPIPS $\downarrow$ 
               & PSNR $\uparrow$ & SSIM $\uparrow$ & LPIPS $\downarrow$ \\ 
\hline
DUSt3R          & 11.66 & 0.47 & 0.63 & 10.72 & 0.46 & 0.67 & 10.81 & 0.40 & 0.68 \\ 
Surf3R-G      & 13.21 & 0.64 & 0.31 & 
                  16.77 & 0.55 & 0.30 & 
                  17.69 & 0.55 & 0.26 \\ 
Surf3R-GD   & \textbf{15.06} & \textbf{0.66} & \textbf{0.26} & 
                  \textbf{17.72} & \textbf{0.61} & \textbf{0.24} & 
                  \textbf{18.08} & \textbf{0.58} & \textbf{0.23} \\ 
\hline
\end{tabular}
}
\caption{\textbf{NVS results on ScanNet++ dataset.} \textbf{Bold} indicates best result. Our method achieves NVS rendering quality comparable with other Gaussian-based methods.
\emph{Surf3R-G}: + Gaussian heads, adds $\mathcal{L}_r$; \emph{Surf3R-GD} (Full model): + D-Normal regularization, adds $\mathcal{L}_s$, $\mathcal{L}_n$ and $\mathcal{L}_{dn}$.
}
\label{tab1}
\vspace{-7pt}
\end{table*}

\subsection{Muliti-view NVS on ScanNet++}
\label{4.3}
As shown in Tab. \ref{tab1}, Surf3R-GD consistently achieves the best novel view synthesis performance across all multi-view configurations on the ScanNet++ dataset. 
With only 4 input views, it outperforms all baselines, achieving a PSNR of 15.06, SSIM of 0.66, and LPIPS of 0.26, demonstrating robust geometry-aware synthesis even under sparse view conditions. 
As the number of input views increases to 12 and 24, our method maintains leading performance, particularly in perceptual quality metrics. 
Notably, the LPIPS score drops to 0.23 at 24 views, outperforming DUSt3R (0.68), indicating more stable view synthesis and sharper geometric details. 
These results underscore the strong generalization capability of our geometry-guided surface reconstruction framework, which not only delivers accurate 3D geometry but also enables high-quality NVS across varying input densities.

\subsection{Ablation Study}
\label{4.4}

\paragraph{View ablations.}
To investigate the impact of input view count on reconstruction quality, we conduct an ablation study by varying the number of input views during inference. As shown in Table~\ref{tab:view}, our method achieves the best performance when using 50 input views, with an F1-score of 41.92. Interestingly, increasing the number of views beyond this point does not lead to further improvements and may even slightly degrade performance. We attribute this to accumulated pose estimation errors from the input point clouds, which become more pronounced as the number of views increases, ultimately affecting the mesh reconstruction quality.

\begin{table}[t]
\centering
\renewcommand{\arraystretch}{1.2}  % 增加行间距
\setlength{\tabcolsep}{5pt}        % 控制列间距
\scalebox{0.8}{
\begin{tabular}{l|ccccc}
\hline
\textbf{Views} & 10 Views & 30 Views & 50 Views & 70 Views & 100 Views \\
\hline
Precision $\uparrow$   & 9.84&  18.41& \textbf{36.66}&  35.42& 34.87\\
Recall $\uparrow$      & 17.57&  30.38& \textbf{49.04}&  47.24& 45.29\\
F1-score $\uparrow$    & 12.29&  22.88& \textbf{41.92}&  40.47& 39.37\\
\hline

\end{tabular}
}
\caption{\textbf{View Ablation Study on Replica dataset.} \textbf{Bold} indicates best result. Performance comparison under different numbers of input views. }
\label{tab:view}
\vspace{-10pt}
\end{table}

\paragraph{Branch and Loss ablations.}
As shown in Tab.~\ref{tab:ablation_loss}, when restricting the network to a single branch (one reference view) degrades the F1-score from 36.66 to 23.24 (Row A). This indicates that, with sparse views, a single reference cannot establish reliable geometric correspondences across wide baselines, resulting in poor reconstructions.
We also verify the effectiveness of different regularization terms on reconstruction quality. 
As shown in Tab.~\ref{tab:ablation_loss}, both components contribute significantly to the reconstruction quality. 
Excluding the scale term (Row B) and the normal term (Row C) results in a notable decline in all metrics.
Notabaly, removing the D-Normal term (Row D) leads to a substantial drop in performance, with the F1-score decreasing from 41.92 to 30.96. This suggests that the D-Normal regularization plays a critical role in encouraging the predicted normals to align with the underlying surface geometry. 
Our full model (Row E) achieves the best performance across all metrics, demonstrating the benefits of both components in producing surface reconstructions.

\begin{table}[ht]
\centering
\renewcommand{\arraystretch}{1.2}
\scalebox{0.8}{
\begin{tabular}{l|ccc}
\hline
\textbf{Ablation Item} & \textbf{Precision $\uparrow$} & \textbf{Recall $\uparrow$} & \textbf{F-score $\uparrow$} \\
\hline
A. w/o Multi-branch   &23.24  &  30.90 & 26.53  \\
B. w/o Scale    & 32.9& 43.20& 37.35\\
C. w/o Normal         & 34.5& 45.80& 39.35 \\
D. w/o D-Normal              & 25.38& 39.69& 30.96\\
E. Full                      & \textbf{36.66}& \textbf{49.04}& \textbf{41.92}\\
\hline
\end{tabular}
}
\caption{\textbf{Branch and Loss Ablation Study on Replica dataset}. \textbf{Bold} indicates best result. Performance with different regularization terms.}
\label{tab:ablation_loss}
\vspace{-10pt}
\end{table}

\section{Conclusion}
In this paper, we propose Surf3R, a novel feed-forward framework for pose-free 3D surface reconstruction from sparse multi-view RGB inputs. Unlike traditional MVS methods that rely heavily on accurate camera calibration and iterative alignment, Surf3R eliminates the need for camera intrinsics or extrinsics by leveraging a cross-view attention mechanism and a multi-branch cross-reference fusion strategy. This enables effective feature propagation across arbitrarily selected views and mitigates the degradation caused by large viewpoint gaps.
Additionally, We introduce a novel Depth-Normal Regularizer grounded in 3D Gaussian representations, which integrates normal estimation into the geometric parameter learning process, yielding more consistent and detailed surfaces.
Extensive experiments on benchmark datasets such as ScanNet++ and Replica demonstrate that Surf3R achieves state-of-the-art performance in surface reconstruction while maintaining strong generalization ability in unseen scenarios.

\bibliography{aaai2026}

\end{document}